\pdfoutput=1
\newif\ifarxiv
\arxivtrue
\ifarxiv\pdfmapfile{+classico.map}\fi
\newif\ifafour
\afourfalse 
\newif\iftypodisclaim 
\typodisclaimtrue

\documentclass[\ifafour a4paper,12pt,\else a5paper,10pt,\fi
onecolumn,oneside,article,
british%
]{memoir}
\newif\ifpublic
\publictrue 

\newcommand*{\firstpublished}{5 March 2018}
\newcommand*{\oggi}{\today}
\newcommand*{\propertitle}{Quantum theory within the probability
  calculus:\\a there-you-go theorem and partially exchangeable models}
\newcommand*{\pdftitle}{Quantum theory within the probability calculus: a there-you-go theorem and partially exchangeable models}
\newcommand*{\headtitle}{Quantum theory within the probability calculus}
\newcommand*{\pdfauthor}{P.G.L.  Porta Mana}
\newcommand*{\headauthor}{\autanet\ Porta Mana}
\newcommand*{\reporthead}{}
\usepackage[T1]{fontenc} 
\input{glyphtounicode} \pdfgentounicode=1
\usepackage[utf8]{inputenx}

\usepackage{textcomp}
\usepackage[normalem]{ulem}
\usepackage{amsmath}
\usepackage{mathtools}
\usepackage{empheq}

\usepackage{framed}
\usepackage{amssymb}
\usepackage{amsxtra}

\usepackage[main=british,french,italian,german,swedish,latin,esperanto]{babel}

\usepackage[autostyle=false,autopunct=false,english=british]{csquotes}
\setquotestyle{american}

\usepackage{amsthm}

\theoremstyle{remark}

\newtheoremstyle{innote}{\parsep}{\parsep}{\footnotesize}{}{}{}{0pt}{}
\theoremstyle{innote}

\usepackage[shortlabels,inline]{enumitem}
\SetEnumitemKey{para}{itemindent=\parindent,leftmargin=0pt,listparindent=\parindent,parsep=0pt,itemsep=\topsep}
\setlist[enumerate,2]{label=\alph*.}
\setlist[enumerate]{leftmargin=\parindent}
\setlist[itemize]{leftmargin=\parindent}
\setlist[description]{leftmargin=\parindent}

\usepackage[babel,theoremfont]{newpxtext}
\usepackage[bigdelims,nosymbolsc
]{newpxmath}
\useosf
\makeatletter
\def\re@DeclareMathSymbol#1#2#3#4{%
    \let#1=\undefined
    \DeclareMathSymbol{#1}{#2}{#3}{#4}}
\re@DeclareMathSymbol{\bigoplusop}{\mathop}{largesymbols}{"4C}
\re@DeclareMathSymbol{\bigotimesop}{\mathop}{largesymbols}{"4E}
\re@DeclareMathSymbol{\sumop}{\mathop}{largesymbols}{"50}
\re@DeclareMathSymbol{\prodop}{\mathop}{largesymbols}{"51}
\re@DeclareMathSymbol{\bigcupop}{\mathop}{largesymbols}{"53}
\re@DeclareMathSymbol{\bigcapop}{\mathop}{largesymbols}{"54}
\re@DeclareMathSymbol{\bigwedgeop}{\mathop}{largesymbols}{"56}
\re@DeclareMathSymbol{\bigveeop}{\mathop}{largesymbols}{"57}
\re@DeclareMathSymbol{\bigtimesop}{\mathop}{largesymbolsPXA}{"10}
\makeatother

\DeclareFontFamily{U}{egreek}{\skewchar\font'177}%
\DeclareFontShape{U}{egreek}{m}{n}{<-6>s*[1]eurm5 <6-8>s*[1]eurm7 <8->s*[1]eurm10}{}%
\DeclareFontShape{U}{egreek}{m}{it}{<->s*[1]eurmo10}{}%
\DeclareFontShape{U}{egreek}{b}{n}{<-6>s*[1]eurb5 <6-8>s*[1]eurb7 <8->s*[1]eurb10}{}%
\DeclareFontShape{U}{egreek}{b}{it}{<->s*[1]eurbo10}{}%
\DeclareSymbolFont{egreeki}{U}{egreek}{m}{it}%
\SetSymbolFont{egreeki}{bold}{U}{egreek}{b}{it}
\DeclareSymbolFont{egreekr}{U}{egreek}{m}{n}%
\SetSymbolFont{egreekr}{bold}{U}{egreek}{b}{n}
\DeclareFontFamily{U}{egreekx}{\skewchar\font'177}
\DeclareFontShape{U}{egreekx}{m}{n}{%
       <-7.5>s*[0.9]euex7%
    <7.5-8.5>s*[0.9]euex8%
    <8.5-9.5>s*[0.9]euex9%
    <9.5->s*[0.9]euex10%
}{}
\DeclareSymbolFont{egreekx}{U}{egreekx}{m}{n}
\DeclareMathSymbol{\sumop}{\mathop}{egreekx}{"50}
\DeclareMathSymbol{\prodop}{\mathop}{egreekx}{"51}
\DeclareMathSymbol{\coprodop}{\mathop}{egreekx}{"60}
\makeatletter
\def\sum{\DOTSI\sumop\slimits@}
\def\prod{\DOTSI\prodop\slimits@}
\def\coprod{\DOTSI\coprodop\slimits@}
\makeatother
\ifarxiv\else\input{undefinegreek.tex}\fi
 \DeclareMathSymbol{\partialup}{\mathalpha}{egreekr}{"40}
 \DeclareMathSymbol{\epsilon}{\mathalpha}{egreeki}{"0F}
 \DeclareMathSymbol{\eta}{\mathalpha}{egreeki}{"11}
 \DeclareMathSymbol{\theta}{\mathalpha}{egreeki}{"12}
 \DeclareMathSymbol{\kappa}{\mathalpha}{egreeki}{"14}
 \DeclareMathSymbol{\lambda}{\mathalpha}{egreeki}{"15}
 \DeclareMathSymbol{\mu}{\mathalpha}{egreeki}{"16}
 \DeclareMathSymbol{\nu}{\mathalpha}{egreeki}{"17}
 \DeclareMathSymbol{\rho}{\mathalpha}{egreeki}{"1A}
 \DeclareMathSymbol{\sigma}{\mathalpha}{egreeki}{"1B}
 \DeclareMathSymbol{\psi}{\mathalpha}{egreeki}{"20}
 \let\varkappa\kappa
 \DeclareMathSymbol{\varDelta}{\mathalpha}{egreeki}{"01}
 \DeclareMathSymbol{\varEpsilon}{\mathalpha}{egreeki}{"45}
 \DeclareMathSymbol{\varIota}{\mathalpha}{egreeki}{"49}
 \DeclareMathSymbol{\varPi}{\mathalpha}{egreeki}{"05}
 \DeclareMathSymbol{\varTau}{\mathalpha}{egreeki}{"54}
 \DeclareMathSymbol{\varOmega}{\mathalpha}{egreeki}{"0A} 
 \DeclareMathSymbol{\deltaup}{\mathalpha}{egreekr}{"0E}
  \DeclareMathSymbol{\piup}{\mathalpha}{egreekr}{"19}

\renewcommand\sfdefault{uop}
\DeclareMathAlphabet{\mathsf}  {T1}{\sfdefault}{m}{sl}
\SetMathAlphabet{\mathsf}{bold}{T1}{\sfdefault}{b}{sl}

\usepackage[scaled=0.84]{DejaVuSansMono}

\usepackage{mathdots}

\usepackage[usenames]{xcolor}
\definecolor{mybluishpurple}{RGB}{51,34,136}
\definecolor{myblue}{RGB}{136,204,238}
\definecolor{mybluishgreen}{RGB}{68,170,153}
\definecolor{mygreen}{RGB}{17,119,51}
\definecolor{mygreenishyellow}{RGB}{153,153,51}
\definecolor{myyellow}{RGB}{221,204,119}
\definecolor{myred}{RGB}{204,102,119}
\definecolor{mypurplishred}{RGB}{136,34,85}
\definecolor{myreddishpurple}{RGB}{170,68,153}
\definecolor{mygrey}{RGB}{221,221,221}
\colorlet{shadecolor}{mygrey}

\usepackage{bm}
\usepackage{microtype}

\usepackage[backend=biber,mcite,
citestyle=authoryear-comp,bibstyle=pglpm-authoryear,autopunct=false,sorting=ny,sortcites=false,natbib=false,maxcitenames=1,mincitenames=1,maxbibnames=8,minbibnames=8,giveninits=true,uniquename=false,uniquelist=false,maxalphanames=1,block=space,hyperref=true,defernumbers=false,useprefix=true,sortupper=false,language=british,parentracker=false]{biblatex}
\DeclareSortingScheme{ny}{\sort{\field{sortname}\field{author}\field{editor}}\sort{\field{year}}}
\DefineBibliographyExtras{british}{\def\finalandcomma{\addcomma}}
\renewcommand*{\finalnamedelim}{\addcomma\space}
\DeclareDelimFormat{multicitedelim}{\addsemicolon\space}
\DeclareDelimFormat{compcitedelim}{\addsemicolon\space}
\DeclareDelimFormat{postnotedelim}{\space}
\ifarxiv\else\fi
\renewcommand{\bibfont}{\footnotesize}
\defbibheading{bibliography}[\bibname]{\section*{#1}\addcontentsline{toc}{section}{#1}
}
\newcommand*{\citep}{\parencites}
\newcommand*{\citey}{\parencites*}
\renewcommand*{\cites}{\parencites}
\providecommand{\href}[2]{#2}

\newcommand*{\amp}{\&}

\newcommand*{\subtitleproc}[1]{}

\usepackage{graphicx}
\usepackage{wrapfig}

\usepackage[hypertexnames=false]{hyperref}
\usepackage[depth=4]{bookmark}
\hypersetup{colorlinks=true,bookmarksnumbered,pdfborder={0 0 0.25},citebordercolor={0.2 0.1333 0.5333},
citecolor=mybluishpurple,linkbordercolor={0.0667 0.4667 0.2},
linkcolor=mypurplishred,urlbordercolor={0.5333 0.1333 0.3333},
urlcolor=mygreen,breaklinks=true,pdftitle={\pdftitle},pdfauthor={\pdfauthor}}

\ifafour\setstocksize{297mm}{210mm}
\else\setstocksize{210mm}{5.5in}
\fi
\settrimmedsize{\stockheight}{\stockwidth}{*}
\setlxvchars[\normalfont] 
\setxlvchars[\normalfont]
\ifafour\settypeblocksize{*}{32pc}{1.618} 
\else\settypeblocksize{*}{26pc}{1.618}
\fi
\setulmargins{*}{*}{1}
\setlrmargins{*}{*}{*}
\setheadfoot{\onelineskip}{2.5\onelineskip}
\setheaderspaces{*}{2\onelineskip}{*}
\setmarginnotes{2ex}{10mm}{0pt}
\checkandfixthelayout[nearest]
\fixpdflayout
\pdfmapline{+dummy-space <dummy-space.pfb}\pdfinterwordspaceon

\newenvironment{acknowledgements}{\section*{Thanks}\addcontentsline{toc}{section}{Thanks}}{\par}
\counterwithout{section}{chapter}
\makeatletter\renewcommand{\appendix}{\par
  \bigskip{\centering
   \interlinepenalty \@M
   \normalfont
   \printchaptertitle{\sffamily\appendixpagename}\par}
  \setcounter{section}{0}%
  \gdef\@chapapp{\appendixname}%
  \gdef\thesection{\@Alph\c@section}%
  \anappendixtrue}\makeatother
\setsecnumformat{\upshape\csname the#1\endcsname\quad}
\setsecheadstyle{\large\bfseries\sffamily%
\raggedright}
\setsubsecheadstyle{\bfseries\sffamily%
\raggedright}
\setaftersubsecskip{-1em}
\setsubsecindent{0pt}
\setparaheadstyle{\bfseries\sffamily%
\raggedright}
\newcommand{\addchap}[1]{\chapter*[#1]{#1}\addcontentsline{toc}{chapter}{#1}}
\newcommand{\addsec}[1]{\section*{#1}\addcontentsline{toc}{section}{#1}}

\copypagestyle{manaart}{plain}
\makeheadrule{manaart}{\headwidth}{0.5\normalrulethickness}
\makeoddhead{manaart}{%
{\footnotesize
\scshape\headauthor}}{}{{\footnotesize\sffamily%
\headtitle}}
\makeoddfoot{manaart}{}{\thepage}{}
\newcommand*\autanet{\includegraphics[height=\heightof{M}]{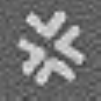}}
\definecolor{mygray}{gray}{0.333}
\iftypodisclaim%
\ifafour\newcommand\addprintnote{\begin{picture}(0,0)%
\put(245,149){\makebox(0,0){\rotatebox{90}{\tiny\color{mygray}\textsf{This
            document is designed for screen reading and
            two-up printing on A4 or Letter paper}}}}%
\end{picture}}
\else\newcommand\addprintnote{\begin{picture}(0,0)%
\put(176,112){\makebox(0,0){\rotatebox{90}{\tiny\color{mygray}\textsf{This
            document is designed for screen reading and
            two-up printing on A4 or Letter paper}}}}%
\end{picture}}\fi
\makeoddfoot{plain}{}{\makebox[0pt]{\thepage}\addprintnote}{}
\else
\makeoddfoot{plain}{}{\makebox[0pt]{\thepage}}{}
\fi
\makeoddhead{plain}{}{}{\footnotesize\reporthead}

\copypagestyle{manainitial}{plain}
\makeheadrule{manainitial}{\headwidth}{0.5\normalrulethickness}
\makeoddhead{manainitial}{%
\footnotesize\sffamily%
\scshape\headauthor}{}{\footnotesize\sffamily%
\headtitle}
\makeoddfoot{manaart}{}{\thepage}{}

\pretitle{\begin{center}\Large\sffamily%
\bfseries}
\posttitle{\bigskip\end{center}}

\makeatletter\newcommand*{\atf}{\includegraphics[
totalheight=\heightof{@}]{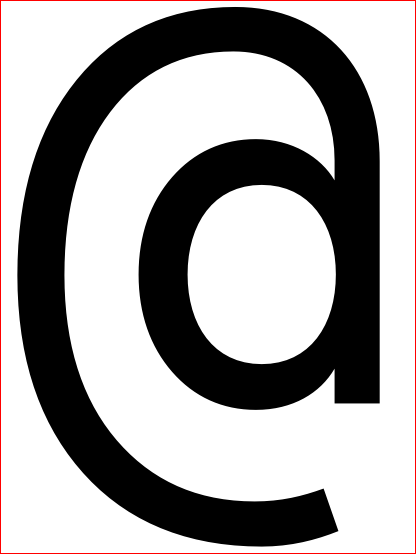}}\makeatother

\providecommand{\epost}[1]{\texttt{\footnotesize\textless#1\textgreater}}
\providecommand{\email}[2]{\href{mailto:#1ZZ@#2 ((remove ZZ))}{#1\protect\atf#2}}

\preauthor{\vspace{-0.5\baselineskip}\begin{center}
\normalsize\sffamily%
\lineskip  0.5em}
\postauthor{\par\end{center}}
\predate{\DTMsetdatestyle{mydate}\begin{center}\footnotesize}
\postdate{\end{center}\vspace{-\medskipamount}}
\usepackage[british]{datetime2}
\DTMnewdatestyle{mydate}%
{
}
\DTMsetdatestyle{mydate}

\setfloatadjustment{figure}{\footnotesize}
\captiondelim{\quad}
\captionnamefont{\footnotesize\sffamily%
}
\captiontitlefont{\footnotesize}
\firmlists*
\midsloppy

\title{\propertitle
}
\author{\ifpublic%
P.G.L. Porta\,Mana%
\else Luca\fi
\quad
\epost{\email{pgl}{portamana.org}}%
}

\date{\firstpublished; updated \oggi}

\newcommand*{\delt}{\deltaup}
\newcommand*{\di}{\mathrm{d}}
\newcommand*{\RR}{\bm{\mathrm{R}}}
\newcommand*{\CC}{\bm{\mathrm{C}}}
\DeclareMathOperator{\tr}{tr}
\DeclareMathOperator{\conv}{conv}

\newcommand*{\suchthat}{\mid}
\renewcommand{\ge}{\geqslant}
\DeclarePairedDelimiter\clcl{[}{]}

\DeclarePairedDelimiter\abs{\lvert}{\rvert}
\DeclarePairedDelimiter\set{\{}{\}}
\newcommand*{\pf}{\mathrm{p}}
\newcommand*{\p}{\mathrm{P}}
\renewcommand*{\|}{\mathpunct{|}}
\newcommand*{\sect}{\S}
\newcommand*{\sects}{\S\S}
\newcommand*{\chap}{ch.}%
\newcommand*{\eqn}{eq.}%

\newcommand*{\etc}{{etc.}}
\newcommand*{\ie}{{i.e.}}
\newcommand*{\eg}{{e.g.}}

\newcommand*{\cf}{{cf.}}
\newcommand*{\etal}{{et al.}}
\newcommand*{\tprod}{\mathop{\textstyle\prod}\nolimits}
\newcommand*{\tsum}{\mathop{\textstyle\sum}\nolimits}

\colorlet{notecolour}{mybluishgreen}
\newcommand*{\puzzle}{\maltese}
\newcommand{\mynote}[1]{ {\color{notecolour}\puzzle\ #1}}

\newcommand*{\yr}{\bm{\rho}}
\newcommand*{\yE}{\bm{\varEpsilon}}
\newcommand*{\yP}{\bm{\varPi}}

\newcommand*{\yS}{S}
\newcommand*{\ySc}{S'}
\newcommand*{\ySt}{\Hat{S}}
\newcommand*{\yM}{M}
\newcommand*{\yMc}{M'}
\newcommand*{\yMl}{M''}
\newcommand*{\yO}{O}
\newcommand*{\yOc}{O''}
\newcommand*{\yT}{T}
\newcommand*{\yTl}{\bm{\varTau}}
\newcommand*{\yR}{R_{\text{personal}}}
\newcommand*{\ys}{\bm{s}}
\newcommand*{\yst}{\Hat{\bm{s}}}

\newcommand*{\yo}{\bm{o}}
\newcommand*{\yq}{q}
\newcommand*{\yqt}{q}
\newcommand*{\yu}{Q}
\newcommand*{\ya}{\lambda}
\DeclarePairedDelimiter\bra{\langle}{\rvert}
\DeclarePairedDelimiter\ket{\lvert}{\rangle}

\firmlists
\begin{document}
\captiondelim{\quad}\captionnamefont{\footnotesize}\captiontitlefont{\footnotesize}
\selectlanguage{british}\frenchspacing

\maketitle
\ifpublic
\abstractrunin
\abslabeldelim{}
\renewcommand*{\abstractname}{}
\setlength{\absleftindent}{0pt}
\setlength{\absrightindent}{0pt}
\setlength{\abstitleskip}{-\absparindent}
\begin{abstract}\labelsep 0pt%
  \noindent \enquote{Ever since the advent of modern quantum mechanics in
    the late 1920's, the idea has been prevalent that the classical laws of
    probability cease, in some sense, to be valid in the new theory.
    \textelp{} The primary object of this presentation is to show that the
    thesis in question is entirely without validity and is the product of a
    confused view of the laws of probability} (Koopman, 1957). The
  secondary objects are: to show that quantum inferences are cases of
  partially exchangeable statistical models with particular prior
  constraints; to wonder about such constraints; and to plead for a 
  dialogue between quantum theory and the theory of exchangeable models.
\end{abstract}\fi

\selectlanguage{british}\frenchspacing


\section{Introduction}
\label{sec:intro}

\begin{quotation}
  \noindent Ever since the advent of modern quantum mechanics in the late
  1920's, the idea has been prevalent that the classical laws of
  probability cease, in some sense, to be valid in the new theory. More or
  less explicit statements to this effect have been made in large number
  and by many of the most eminent workers in the new physics \textelp{}.
  Some authors have even gone farther and stated that the formal structure
  of logic must be altered to conform to the terms of reference of quantum
  physics \textelp{}.

  Such a thesis is surprising, to say the least, to anyone holding more or
  less conventional views regarding the positions of logic, probability,
  and experimental science: many of us have been apt -- perhaps too naively
  -- to assume that experiments can lead to conclusions only when worked up
  by means of logic and probability, whose laws seem to be on a different
  level from those of physical science.

  The primary object of this presentation is to show that the thesis in
  question is entirely without validity and is the product of a confused
  view of the laws of probability.
\sourceatright{(B. O. Koopman, \cite*{koopman1957})}
\end{quotation}

Koopman's lucid presentation stands today as it did sixty years ago.
Perhaps it can be made even clearer if we consider quantum systems with a
finite number of energy levels -- qubits, qutrits, \etc, very important in
today's chase for quantum computers \citep{nielsenetal2000_r2010} -- and if
adopt the so-called \enquote{operational approach} to quantum theory, which
can be glimpsed in Koopman's presentation itself.

\section{Operational approach}
\label{sec:operational_approach}

A pseudohistorical presentation:

After decades of inconclusive debates about what quantum systems
\enquote{really} are, the measurement problem, Schödinger's cats, and
similar questions, the operational approach
\cites[see \eg:][]{ludwig1954_t1983,segal1959,mielnik1968,mielnik1969,lamb1969,daviesetal1970,foulisetal1972,randalletal1973,randalletal1979,wright1978,haag1982,kraus1983,wootters1986,buschetal1995b}[for
more recent
elaborations:][]{hardy2001,portamana2003b,portamana2004b,barnumetal2006,barrett2007,harriganetal2007b}
emerged as a way to sidestep or postpone answering them and get (blindly?)
on with experiments and technology.

\medskip

The starting points of this approach are these:
\begin{enumerate}[label=\Roman*.,ref=\Roman*]
\item\label{item:verbal}\enquote{all well-defined experimental evidence, even if it
    cannot be analysed in terms of classical physics, must be expressed in
    ordinary language making use of common logic \textelp{}. This is a
    simple logical demand, since the word \guillemotleft
    experiment\guillemotright\ can in essence only be used in referring to a
    situation where we can tell others what we have done and what we have
    learned} \citep{bohr1948}.
\item\label{item:prep_meas_outc} This verbalization of every experiment is
  conveniently divided into three parts: the descriptions of a
  \emph{preparation}, of a \emph{measurement}, and of several possible
  \emph{outcomes}; each represented by a proposition: $\yS$, $\yM$,
  $\yO_i$. The outcomes, implicit in the description of the measurement,
  can be probabilistically predicted; including deterministic,
  unit-probability predictions. For each quantum system we have a set of
  possible preparations and a set of possible measurements; elements from
  the two sets can be freely combined, at least in principle.
\item\label{item:qm_associations} To a preparation $\yS$ we can associate a
  unit-trace, positive-definite Hermitean matrix $\yr$ usually called
  \emph{density matrix}; and to the outcomes $\set{\yO_i}$ of a measurement
  $\yM$, a set of positive-definite Hermitean matrices $\set{\yE_i}$
  summing up to the identity matrix; this set is called a
  \emph{positive-operator-valued measure}. The dimension of these matrices depends on the system.
\item\label{item:trace_formula} The probability of obtaining outcome
  $\yO_i$ of the measurement $\yM$ when the preparation is $\yS$ is given
  by
  \begin{equation}
    \label{eq:trace_formula}
    \p(\yO_i \| \yM \land \yS) = \tr\yE_i\yr,
  \end{equation}
  usually called the \emph{trace formula}. The properties of the
  matrices guarantee that the probability distribution for the outcomes
  $\set{\yO_i}$ is non-negative and normalized.
\end{enumerate}
In the verbalization of an experiment we can also include the description
of a \emph{transformation}, possibly parameterized by time; this is where
Schrödinger's equation appears. Transformations are briefly discussed in
appendix~\ref{sec:transformations}. For the moment let's keep the
description of this operational approach to a minimum.
Appendix~\ref{sec:traditional_qt} shows how this approach comprises the
old-fashioned quantum formalism with Hermitean operators \amp\ Co.

The operational approach favours the view of probability as an extension of
the propositional truth calculus
\citep{keynes1921_r1957,johnson1924,ramsey1926,cox1946,polya1949,jaynes1994_r2003,hailperin1996,hailperin2011,tereninetal2015_r2017}.
Sure, we can translate all this in terms of \enquote{random variables}
about physical quantities, but the verbal and propositional character of
this approach is fundamental. It works because quantum physicists usually
agree on the coarsest, protocol-like verbal description of an experiment,
even if they may disagree on what is \enquote{really} going on
microscopically. They agree on how to divide the experiment into
preparation, measurement, and outcomes. They agree on which density
matrices and positive-operator-valued measures to associate with those
divisions. Each physicist can add his or her own personal interpretation
$\yR$ of what is \enquote{really} going on, but it becomes irrelevant when
the coarse preparation is specified; we could write this irrelevance as
\begin{equation}
  \label{eq:metaphysical_irrelevance}
  \p(\yO_i \| \yM \land \yS \land \yR) = \p(\yO_i \| \yM \land \yS).
\end{equation}

The operational approach will thus still be valid if we'll eventually agree
on a microscopic interpretation of quantum phenomena.

We shall now find additional reasons for the propositional view of
probability in this approach.

\section{Convexity of preparations}
\label{sec:convexity}

The operational approach was accompanied by several developments in the
mathematical formalism of quantum theory (\eg, the use of
positive-operator-valued measures), recruiting from subjects like
$C^*$-algebras, lattice theory, convex spaces. The latter I find most
insightful.

The set of positive-definite Hermitean matrices associated with
preparations and measurement outcomes in
points~\ref{item:qm_associations}--\ref{item:trace_formula} above can be
seen as a subset of a \emph{real} vector space of dimension $n^2$, where
$n$ is the dimension of these matrices. The trace product in
\eqn~\eqref{eq:trace_formula} is just a scalar multiplication of such
vectors (or better, the contraction of a vector and a dual 1-form, without
scalar products). This means that we can associate a \emph{real}-valued
vector $\ys$ with each preparation, and a set of real-valued vectors
$\set{\yo_i}$ with each set of measurement outcomes, and the trace
formula~\eqref{eq:trace_formula} becomes
\begin{equation}
  \label{eq:vector_formula}
  \p(\yO_i \| \yM \land \yS) = \yo_i\cdot\ys.
\end{equation}
A brilliant paper by Hardy \citey{hardy2001}, foreshadowed by Wootters
\citey{wootters1986}, showed that this formula is true for any physical
theory: quantum, classical, or otherwise. In fact, it holds for any
collection of three kinds of propositions satisfying
points~\ref{item:verbal}--\ref{item:prep_meas_outc}, whether they be about
physical theories or not \citep{portamana2003b,portamana2004b}.

The sets of vectors $\set{\ys}$ and $\set{\set{\yo_i}}$ satisfy constraints
that guarantee
\setlength{\intextsep}{0ex}%
\begin{wrapfigure}{r}{0.45\linewidth}
 \centering \includegraphics[width=\linewidth]{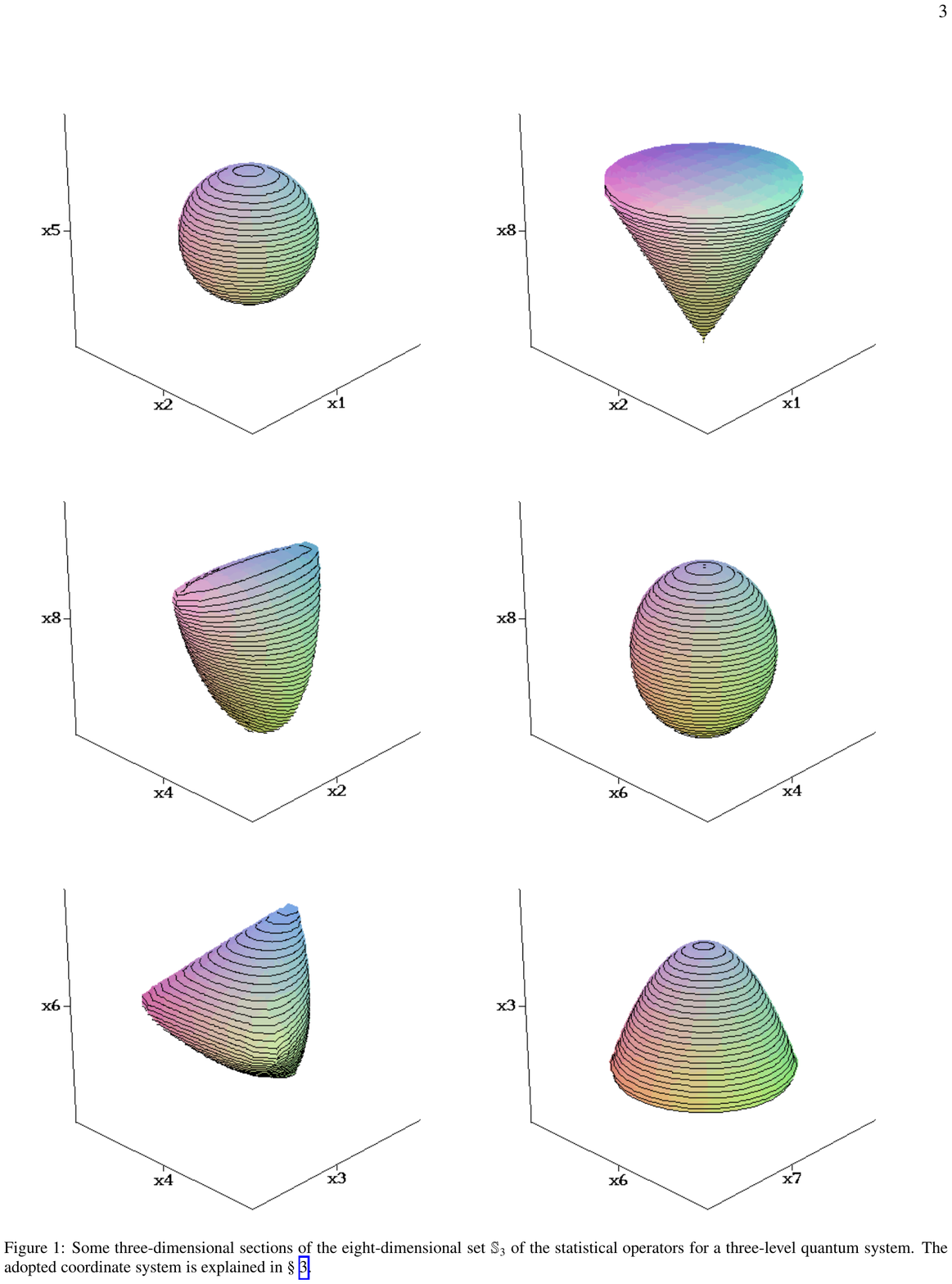}
 \\ \footnotesize 3D sections of the convex hull of $\CC\mathrm{P}^{2}$\\
\citep{maanssonetal2006}
\end{wrapfigure}
the positivity and normalization of the probabilities; these constraints
say that these sets are convex spaces. Classical and quantum systems differ in
the convex properties of their sets of vectors $\set{\ys}$; let's call
these \emph{states}, and let's call \emph{extremal} the preparations that
are represented by extremal states of these convex sets. The set of states
of a classical system is a simplex; that of a quantum system is the convex
hull of complex projective space $\CC\mathrm{P}^{n-1}$
\citep{bengtssonetal2006}. See figure on the right for an example with
$n=3$ (qutrit). Because of these differing convex structures, for a
classical system there is always a measurement that allows us to infer with
probability $1$ which of two extremal preparations was made:
\begin{multline}
  \label{eq:extremal_classical}
\text{there is $\yM$ such that}\quad  \p(\yS \| \yM \land \yO_i) = 0\text{ or }1,\\
  \yS\in\set{\text{extremal preparations}};
\end{multline}
whereas for a quantum system this is possible for particular extremal
preparations only. These different behaviours under inference aren't
foreign to the probability calculus, however. Kirkpatrick
\citey{kirkpatrick2001_r2003,kirkpatrick2002_r2003} showed that analogous
inferential characteristics appear in some games with cards, for example;
similar examples are easily constructed with urn-drawing
\citep[\sect~IV]{portamana2003_r2004}.

Still today nobody knows why the sets of states of quantum systems have a
projective-space convex structure. This is the \enquote{\emph{only}
  mystery} \citep[\sect~1.1]{feynmanetal1965} of quantum theory.

\medskip

In an experiment with a physical system -- quantum or otherwise -- we can
imagine a state of knowledge $\ySc$ where we are unsure about which of two
preparations $\yS_1$, $\yS_2$ was made, with corresponding probabilities:
\begin{equation}
  \label{eq:uncertain_preps}
  \p(\yS_1 \| \ySc) = \yq_1, \quad
  \p(\yS_2 \| \ySc) = \yq_2,
  \qquad \yq_1+\yq_2=1.
\end{equation}
In this state of knowledge our predictions for any measurement outcome will
be, by the probability calculus,
\begin{multline}
  \label{eq:prediction_uncertain_prep}
    \p(\yO_i \| \yM \land \ySc) = 
    \p(\yO_i \| \yM \land \yS_1)\; \p(\yS_1 \| \ySc) +
    \p(\yO_i \| \yM \land \yS_2)\; \p(\yS_2 \| \ySc) ={}\\ 
    (\yo_i \cdot \ys_1)\, \yq_1 +
    (\yo_i \cdot \ys_2)\, \yq_2  =
    \yo_i \cdot (\yq_1\ys_1 + \yq_2\ys_2).
\end{multline}
The first equality tacitly uses some logical-independence assumptions that
are quite natural in an experimental setup; \eg, the choice of measurement
doesn't tell us anything about the preparation.

The last equality says that we can associate the vector
$\yq_1\ys_1 + \yq_2\ys_2$, a convex combination of the states $\ys_1$ and
$\ys_2$, with the state of knowledge $\ySc$. This state of knowledge,
usually called a \emph{mixture}, can therefore be considered a citizen of
the set of preparations.

It is natural to assume that states of knowledge like $\ySc$ exist with all
possible values of the distributions $(\yq_1, \yq_2)$, and also involving
more than two preparations. This assumption implies that the \enquote{domain of
  discourse} for our system, even if it initially has only a finite number
of preparations $\set{\yS}$ represented by states $\set{\ys}$, can always
be extended to an infinite number of preparations, corresponding to the
convex hull of $\set{\ys}$.

\section{Lattice structure of measurements}
\label{sec:lattice}

We can also consider two kinds of state of knowledge involving
measurements and outcomes \citep[\cf][]{peresetal1998}.

The first, as with preparations, is a state of knowledge $\yMc$ where we are
unsure whether measurement $\yM_1$ or $\yM_2$ was made, with probabilities
\begin{equation}
  \label{eq:uncertain_meas}
  \p(\yM_1 \| \yMc) = \yq_1, \quad
  \p(\yM_2 \| \yMc) = \yq_2,
  \qquad \yq_1+\yq_2=1.
\end{equation}
The set of outcomes allowed by this state of knowledge is
$\set{\yO_{1i}} \cup \set{\yO_{2j}}$, and we have
\begin{multline}
  \label{eq:prediction_uncertain_meas}
    \p(\yO_{1i} \| \yMc \land \yS) = 
    \p(\yO_{1i} \| \yM_1 \land \yS)\; \p(\yM_1 \| \yMc) +
    \p(\yO_{1i} \| \yM_2 \land \yS)\; \p(\yM_2 \| \yMc) ={}\\
    (\yo_{1i} \cdot \ys)\, \yq_1 + 0 =
    (\yq_1\yo_{1i}) \cdot \ys,
\end{multline}
and analogously for $\yO_{2j}$, assuming that the two original sets of
outcomes are mutually exclusive. The state of knowledge $\yMc$, usually
called a \emph{mixture}, can thus be considered a measurement, associated
with the vectors $\set{\yq_1\yo_{1i}} \cup \set{\yq_2\yo_{2j}}$.

The second is a state of knowledge $\yMl$ in which we are not interested in
the outcomes $\set{\yO_i}$ of a particular experiment $\yM$, but in other
events $\set{\yOc_j}$ which we can probabilistically infer from those
outcomes, with
\begin{equation}
  \label{eq:uncertain_outcomes}
  \p(\yOc_j \| \yO_i\land\yMl) = \yu_{ji}, 
  \qquad \tsum_j\yu_{ji}=1.
\end{equation}
Then
\begin{multline}
  \label{eq:prediction_coarsen_meas}
  \p(\yOc_j \| \yMl \land \yS) = 
  \sum_i \p(\yOc_j \| \yO_i\land\yMl)\; \p(\yO_i\| \yM \land \yS) ={}\\
  \sum_i\yu_{ji}\; (\yo_i\cdot \ys) =
  \bigl(\tsum_i\yu_{ji} \yo_i\bigr)\cdot \ys,
\end{multline}
and we can consider $\yMl$ also as a measurement, associated with the
vectors $\set{\tsum_i\yu_{ji}\yo_i}$. We can call it a \emph{dither},
but it includes \emph{coarsenings}, \ie\ situations where we aren't
interested in distinguishing several outcomes; for example considering the
set $\set{\yO_1\lor\yO_2, \yO_3}$ instead of $\set{\yO_1, \yO_2, \yO_3}$.

Mixtures and dithers can also be combined. They make the set of measurement
outcomes into something more than a convex set: it is a set of lattices
that can be combined in the two ways just described.

\section{Conclusion}
\label{sec:conclusion}

\enquote{Conclusion?} you might ask. \enquote{Wasn't all the above just the
  preamble to a Proof that the standard probability calculus suffices for
  quantum theory? Where's the proof?} The proof was under your eyes as you
were reading. The mathematical formalism just presented covers that of
quantum systems with a finite number of energy levels, and with some
topological care can be extended to cover systems with continuous energy
levels, the Schrödinger equation, and even quantum field theory as infinite
limits. This mathematical formalism is so general that it can also describe
exotic systems that have neither classical nor quantum inferential traits.
Did we have to generalize the ordinary probability calculus for this
formalism? did we use any exotic probability theory? No. There you go.

Neither complex-valued probabilities nor lattices of $\sigma$-algebras have
been necessary. In fact, what we've done looks simply like a special
application of the probability calculus. The only special feature is the
vector-product formula~\eqref{eq:vector_formula}, expressing some
probabilities as the results of vector products. But the probability
calculus doesn't care where the numerical values of its probabilities come
from, as long as they don't break its basic rules. And as we've seen, its
rules are not broken when dealing with quantum experiments. Moreover, the
vector-product formula doesn't have any physical significance: it appears
when our domain of discourse involves three kinds of propositions logically
related in a particular way \citep{portamana2003b,portamana2004b}, but the
propositions could be, say, about Donald Duck or parallel universes or
other such things.

Only the particular convex structure of the preparations has physical
significance, and it's experimentally observed. But the usual
probability calculus can accommodate every convex structure, including the
one peculiar to quantum theory.

\medskip

The notion that quantum systems require lattices of $\sigma$-algebras --
and therefore a generalization of the probability calculus -- arises when
we ignore the specification of the measurement in the conditional of the
probabilities $\p(\yO_i \| \yM \land \yS)$. We end up with several sets
$\varOmega_{\yM}$ and their $\sigma$-algebras, one for each $\yM$.

But this mathematical move does not make much sense, for at least three + one
reasons.

First, in real applications we often need to consider uncertainties about
measurement procedures
\citep[\chap~7]{leonhardt1997,demuynck2002b}{zimanetal2004_r2006,darianoetal2004}.
To do so we must use the mixing, dithering, coarsening
formulae~\eqref{eq:uncertain_meas}--\eqref{eq:prediction_coarsen_meas}
within the standard probability calculus. In other words, we must gather
the flock of separate sets $\varOmega_{\yM}$ and $\sigma$-algebras
back together, as subsets and subalgebras of one set only.

Second, the lattice structure of these $\sigma$-algebras reflects the
operations of measurement mixing and dithering described in
\sect~\ref{sec:lattice}, operations clearly arising from conditionalization
within one $\sigma$-algebra only.

Third, the game of wearing blinkers in order to see seemingly separate
$\sigma$-algebras can be played in non-quantal, everyday contexts, like
card or urn-drawing games
\citep{kirkpatrick2001_r2003,kirkpatrick2002_r2003}[\sect~IV]{portamana2003_r2004}.
Are these also \enquote{quantal}?

What's worse, this mathematical move inhibits an already non-existent
dialogue between quantum theory and the theory of statistical models based
on exchangeability, sufficiency, symmetry \citep[for a glimpse
see][\chap~4]{bernardoetal1994_r2000} that has been flourishing in
probability and statistics since the 1930s, with many brilliant results and
papers -- \eg\ those by Koopman \amp\ Pitman
\citep{koopman1936,pitman1936,darmois1935}, Diaconis \amp\ Freedman
\citep{freedman1962,freedman1962b,diaconis1977,diaconis1988,diaconis1992,diaconisetal1980,diaconisetal1980b,diaconisetal1980c,diaconisetal1981,diaconisetal1987,diaconisetal1988b,diaconisetal1990},
Martin-Löf \citey{martinloef1974}, Lauritzen
\citey{lauritzen1974,lauritzen1974b,lauritzen1982_r1988,lauritzen1984,lauritzen2007},
Ressel \citey{ressel1983_r1985}, Aldous
\citey{aldous1981,aldous1982,aldous1985,aldous2010}, Kallenberg
\citey{kallenberg1989,kallenberg2005}, Cifarelli, Regazzini, Fortini,
\etal\
\citep{cifarellietal1979,cifarellietal1980,cifarellietal1981,cifarellietal1982,regazzini1996,fortinietal2000,fortinietal2002,fortinietal2012,fortinietal2014},
to name very few besides those by de~Finetti
\citey{definetti1930,definetti1937,definetti1938}, already known in the
quantum literature. See Dawid's review \citey{dawid2013} for a small
glimpse. Such a dialogue would surely benefit both disciplines, as I hope
the ideas presented in the next section show.

\section{Quantum theory as a partially exchangeable model}
\label{sec:quantum_exchangeability}



Many inferences in physics are instances of infinitely
exchangeable statistical models
\citep[\sects~4.2--3]{bernardoetal1994_r2000}:
\begin{equation}
  \label{eq:param_exch_model}
  \pf(D^{(1)}, D^{(2)}, \dotsc \| H) =
  \int \Bigl[\prod_i\pf(D^{(i)} \| \theta, H)\Bigr] \,
  \pf(\theta \| H) \, \di\theta,
\end{equation}
where $D^{(i)}\in\set{\yO_j}$ are observed outcomes of a set of experiments
$(1)$, $(2)$, \ldots\ made in identical conditions and each
$\pf(D^{(i)} \| \theta, H)$ is a categorical distribution (\ie\ one-trial
multinomial). This expression can be interpreted as a mixture of product
probabilities $\pf(D^{(i)} \| \theta, H)$ indexed by the vector parameter
$\theta$, weighted by the distribution $\pf(\theta \| H)$. The integration
is defined over a simplex, but the distribution $\pf(\theta \| H)$ can
effectively restrict it to a subset thereof. The distribution on the left
side is usually called the \emph{predictive} distribution.

The integral formula above results automatically when we assume that the
joint probability of any number of outcomes is invariant under their
permutations, no matter how many outcomes we consider. This assumption is
called \emph{infinite exchangeability}, and this result is de~Finetti's
representation theorem \parentext{\cite*{definetti1930};
  \cite{hewittetal1955}}. The theorem leaves undetermined the distribution
$\pf(\theta \| H)$ only, usually called the \emph{prior}. All infinitely
exchangeable distributions over the outcomes are in one-one correspondence
with all distributions $\pf(\theta \| H)$.

The exchangeability assumption can in turn be motivated by the identical
condition in which the experiments were made. The remarkable part of this
representation is that it automatically introduces mathematical objects
analogous to the statistical states (Liouville distributions) $\set{\ys}$
of a discrete classical system: $\set{\theta} \equiv \set{\ys}$. We can
interpret it as saying that each experiment was independently made with the
same -- but unknown -- preparation $\yS$. Hence the integral, with the
prior $\pf(\theta \| H) \equiv \pf(\ys \| H)$ representing our knowledge
$H$ about the preparation. The features of this prior, like its support and
maxima, may thus be motivated by physical laws or constraints.

Many authors \parentext{see the list at the end of
  \sect~\ref{sec:conclusion}} later proved various generalizations of this
representation theorem, extending it to predictive distributions invariant
under other symmetry groups, or possessing sufficient statistics.

\medskip

Inference for quantum systems does not quite fit within the simple
statistical model above, however. As we saw in the previous sections,
quantum systems allow for a set $\set{\yM}$ of distinct measurements that
cannot be obtained by marginalization from one another. Inferences for such
systems therefore require that the conditional of the predictive
distribution above specify which measurements $\yM^{(i)} \in \set{\yM}$ are
performed, as shown in the probabilities of the previous sections. If we
again interpret these experiments as independently made with the same but
unknown preparation, we arrive \citep{portamanaetal2006} at the expression
\begin{multline}
  \label{eq:unknown_q_state}
  \pf(D^{(1)}, D^{(2)}, \dotsc \| \yM^{(1)}, \yM^{(2)}, \dotsc, H) ={}\\
  \int\limits_{\mathclap{\conv\CC\mathrm{P}^{n-1}}} \Bigl[\tprod_i\pf(D^{(i)} \| \yM^{(i)}, \ys,  H)\Bigr] \,
  \pf(\ys \| H) \, \di\ys.
\end{multline}
The integration is over the convex hull of complex projective space
$\CC\mathrm{P}^{n-1}$ \citep{bengtssonetal2006}, as explained in
\sect~\ref{sec:convexity}, where $n$ is the number of quantum states
completely distinguishable with a single measurement.

The expression above is a particular case of a \emph{partially exchangeable
  model}
\cites[\sects~4.6]{bernardoetal1994_r2000}[\chap~5]{gelmanetal1995_r2014}.
The assumption of \emph{partial} exchangeability states that the predictive
distribution is invariant under permutations of outcomes of the same kind
of measurement, but not across different kinds of measurement. This makes
sense also because different measurements may have different numbers of
outcomes. For examples of when and why this kind of assumption arises see
\textcite[\chap~4]{bernardoetal1994_r2000}.

The assumption of partial exchangeability leads to a representation theorem
too \citep{definetti1938,bruno1964,diaconisetal1980b}, of the form
\begin{multline}
  \label{eq:partial_exch}
  \pf(D^{(1)}, D^{(2)}, \dotsc \| \yM^{(1)}, \yM^{(2)}, \dotsc, H) ={}\\
  \int \Bigl[\prod_i\pf(D^{(i)} \| \yM^{(i)}, \eta_{\yM^{(i)}},  H)\Bigr] \,
  \pf[(\eta_{\yM}) \| H] \, \prod_{\yM}\di\eta_{\yM},
\end{multline}
where each outcome $D^{(i)}$ belongs to the set of possible outcomes of
measurement $\yM^{(i)}$, these measurements belong to the system's set of
possible measurements, $\yM^{(i)} \in \set{\yM}$, and each
$\pf(D^{(i)} \| \yM^{(i)}, \eta_{\yM^{(i)}}, H)$ is a categorical
distribution. Just like the expression~\eqref{eq:param_exch_model} for
infinite exchangeability, also this expression is a mixture of products of
distributions indexed by parameters $(\eta_{\yM})$, one for each kind of
measurement, weighted by the prior
$\pf[(\eta_{\yM}) \| H]$. The integration is
defined over the \emph{Cartesian product of simplices}
$\prod_{\yM}\set{(\eta_{\yM})}$, but the prior can effectively restrict it to
a subset thereof. There is again a one-one correspondence between all
partially exchangeable predictive distributions (left side) and all priors.
The expression~\eqref{eq:param_exch_model} for infinite exchangeability is
a special case of the one above when $\set{\yM}$ comprises only one kind of
measurement.

This generalized representation theorem is remarkable because it also
automatically introduces a mathematical object, the parameter space
$\set{(\eta_{\yM})}$, which is similar to a space of states. We can
interpret it as saying that each experiment was independently made with the
same -- but unknown -- preparation $(\eta_{\yM})$. Also in this case the
prior $\pf[(\eta_{\yM}) \| H]$ expresses our
knowledge $H$ about the possible preparations, and its functional features
can be motivated by physical laws or constraints.

\medskip

We said that the inferential formula~\eqref{eq:unknown_q_state} for quantum
systems is a particular case of the expression~\eqref{eq:partial_exch} for
partial exchangeability. Let's see what additional features make it a
particular case. First we have to identify the states $\ys$ in the former
with the parameters $(\eta_{\yM})$ in the latter. Then we see that it is a
particular case because \emph{the support of the prior
  $\pf[(\eta_{\yM}) \| H]$ is restricted to a
  particular lower-dimensional convex subset of the Cartesian product}: the
convex hull of a complex projective space. This restriction reflects the
statistical properties of quantum systems, and is remarkably strong,
because it reduces the support of the prior distribution from an
infinite-dimensional manifold to a finite-dimensional one, for example from
the function space
$\set{f \suchthat f \colon \RR\mathrm{P}^{2} \to \clcl{0,1}}$ to the
three-dimensional ball $\CC\mathrm{P}^{1}$ in the case of a qubit.

The full partially exchangeable model~\eqref{eq:partial_exch} allows for
more general cases: less constrained than the quantum case, \eg\ where each
measurement has an outcome having probability $1$; and more constrained
than the quantum case, \eg\ where no measurement can ever have a sure
outcome.

For a quantum system, the a priori restriction on the prior of the
partially exchangeable model reflects our empirical observation of
measurement-outcome constraints typical of such systems: uncertainty
relations, \etc. For example, if preparing an electron spin in a particular
way we have probability $1$ of obtaining $+z$ in a measurement along the
$z$ direction, then with the same preparation we cannot have probability
$1$ or $0$ of obtaining $+x$ in a measurement along the $x$ direction. We
still don't know why such constraints exist. The mathematical formalism of
quantum theory expresses and enforces these constraints, but doesn't
explain why they exist either; just like the equation $\di S/\di t\ge0$
\citep{truesdell1969_r1984,owen1984} expresses and enforces the empirically
found second law of thermodynamics, but doesn't explain why it must be so.

\medskip

The exchangeability representation theorems~\eqref{eq:param_exch_model}
and~\eqref{eq:partial_exch} are welcome by many scientists, including yours
faithfully, because they pull the notion of \emph{state} out of the hat,
thus also demoting it to a secondary, in principle avoidable notion. And
they do so by promoting the notion of repeated, reproducible experiments,
which science indeed hinges on. This point of view has been forcefully
promoted by some probabilists in recent years \citep[see
\eg][]{cifarellietal1982,regazzini1996,fortinietal2000}.

The partially exchangeable model, including quantum inferences as special
cases, thus demotes quantum states too. But it does so without
de-emphasizing the physical and inferential properties characteristic of
quantum systems, properties still reflected in the peculiar constraints of
the model's prior. And moreover it emphasizes the necessity of considering
\emph{several} distinct measurements when dealing with quantum systems.
These emphases should be contrasted with the features of the
\enquote{quantum} exchangeability representation theorem by Hudson \etal\
\citep{hudsonetal1976,hudson1981}, neatened by Caves \etal\
\citey{cavesetal2002}. This representation is surely useful in applications
\citep{vanenketal2002,fuchsetal2004b}. But it's tailor-made for quantum
systems and therefore partially veils their peculiarity, if only by not
openly showing the whole infinite-dimensional space of \enquote{unknown
  states} allowed by the full model; and it also veils the fact that quantum
inferences need an assumption of \emph{partial} exchangeability.

\medskip

In the theory of exchangeable models it is known
\cites{lauritzen1982_r1988,ressel1983_r1985,diaconis1988,diaconis1992,kallenberg2005,dawid2013}
that the symmetries of a predictive distribution imply a particular form of
its \emph{likelihood} and the space of parameters -- hence the support of
the prior -- in the integral representation. This leads to interesting
questions\ldots

\begin{itemize}[\ldots]
\item for quantum theory: What kinds of symmetries or physical laws could
  cause the particular restrictions on the prior of the partially
  exchangeable model? The quantum literature offers studies of possible
  such symmetries \citep{wigner1931_t1959,haag1992_r1996,holevo1980_t2011},
  but their discussion disregards partially exchangeable models.

\item for the theory of exchangeable models: In which other contexts can or
  do analogous restrictions on the prior appear? Say, finance? biology? In
  which other contexts could they fruitfully be employed?

\end{itemize}

\smallskip

\noindent\enquote{Consent with both theories that they may enjoy each
  other: it shall be to our good} \citep[scene~V.II]{shakespeare1623}.

\setlength{\intextsep}{0.5ex}

\ifpublic
\begin{acknowledgements}
  \ldots to H. Barnum, A. Terenin, D. Draper for unintentionally rekindling
  my interest in these matters. To Mari \amp\ Miri for continuous
  encouragement and affection. To Buster Keaton for filling life with awe.
  To the developers and maintainers of \LaTeX, Emacs, AUC\TeX, Open Science
  Framework, biorXiv, Hal archives, Python, Inkscape, Sci-Hub for making a
  free and unfiltered scientific exchange possible.
\end{acknowledgements}
\fi

\appendix

\vspace{-\medskipamount}

\section{Transformations}
\label{sec:transformations}

In the operational approach, the \emph{transformation} of a preparation is
some knowledge $\yT$, possibly dependent on a parameter like time (hence a
collection of such propositions), that allows us to infer one preparation
from another, possibly at different times, and possibly across different
physical systems:
\begin{equation}
  \label{eq:transformation}
  \p(\ySt \| \yT \land \yS) =\yqt.
\end{equation}
This notion can be shown to subsume the usual notions of deterministic
evolution, stochastic evolution, and collapse, and can be naturally
combined with all the probabilistic formulae we have seen so far. Every
transformation can be associated with a linear map acting on the state
vectors: $\yst = \yTl \ys$. We can also consider mixtures of
transformations, and so on.

It's important to note that Schrödinger's equation describes the
time-dependence \emph{of the linear map} $\yTl$ associated with a particular
transformation $\yT$ -- not of the probability \eqref{eq:transformation}
itself.

\section{Traditional quantum theory from the operational approach}
\label{sec:traditional_qt}

In the space of such matrices of positive-definite Hermitean matrices of
dimension $n$ we can always find $n$ orthogonal projectors $\set{\yP_i}$
such that $\yP_i\yP_j = \delt_{ij}\yP_i$. Such projectors have unit trace;
they can thus represent the density matrix associated with a preparation.
They can also be written as $\ket{\psi_i}\bra{\psi_i}$, where
$\bra{\psi_i}$ is a unit complex vector and $\ket{\psi_i}$ its dual. The
sum of orthogonal projectors is the identity matrix; a set of orthogonal
projectors can thus be associated with the outcomes of a measurement, too.

When we consider a preparation and a measurement outcome associated with
orthogonal projectors $\yE_k$, $\yE_i$, the trace formula becomes the famous
\begin{equation}
  \label{eq:born_rule}
  \p(\yO_i \| \yM \land \yS) =
  \tr\yE_i\yE_k \equiv \abs{\bra{\psi_i}\ket{\psi_k}}^2.
\end{equation}

If the outcomes $\set{\yO_i}$ of a measurement associated with orthogonal
projectors are numerical values $\set{\ya_i}$, the expected value is
another famous formula:
\begin{equation}
  \label{eq:exp_value}
  \sum_i \ya_i \p(\yO_i \| \yM \land \yS) =
  \sum_i\ya_i\tr\yE_i\yr =
  \bra{\psi_k}\bigl[ \tsum_i\ya_i\ket{\psi_i}\bra{\psi_i}\bigr]\ket{\psi_k},
\end{equation}
where the expression in brackets is a Hermitean operator with real
spectrum.

Orthogonal projector matrices are associated with preparations and
measurements which jointly lead to completely certain outcomes, in the
sense of \eqn~\eqref{eq:extremal_classical}. The more general density
matrices and positive-operator-valued measures were introduced to describe
experimental situations in which noise sources make the preparation
uncertain, and interaction with other systems during measurement can lead
to noise in the outcomes or even to their proliferation
\cites{buschetal1989c}[\chap~7]{demuynck2002b}. The more general
measurements associated with positive-operator-valued measures also include
\emph{simultaneous} measurements of conjugate quantities like position and
momentum \citep{arthursetal1965,buschetal1984,appleby1998}, which are
routine in fields like quantum optics \citep[\chap~6]{leonhardt1997}.


\defbibnote{prenote}{{\footnotesize (\enquote{de $X$} is listed under D,
    \enquote{van $X$} under V, and so on, regardless of national
    conventions.)\par}}

\printbibliography[prenote=prenote
]

\end{document}